# Magneto-resistive memory in ferromagnetic (Ga,Mn)As nanostructures

T. Figielski*,[1], T. Wosinski[1], A. Morawski[1], A. Makosa[1], J. Wrobel[1] and J. Sadowski[1,2,3]

[1]*Institute of Physics, Polish Academy of Sciences, 02-668 Warszawa, Poland*
[2]*MAX-Lab, Lund University, 221 00 Lund, Sweden*
[3]*Groupe d'Etude des Semiconducteurs CC074, Université Montpellier II, Montpellier, France*

We show a novel magneto-resistive effect that appears in lithographically shaped, three-arm nanostructure, fabricated from ferromagnetic (Ga,Mn)As layers. The effect, related to a rearrangement of magnetic domain walls between different pairs of arms in the structure, reveals as a dependence of zero-field resistance on the direction of previously applied magnetic field. This effect could allow designing devices with unique switching and memory properties.

*Electronic mail: figiel@ifpan.edu.pl

A great effort has been made of late decade to develop a basis for novel spin electronics, in which the electron spin performs functions like those the electron charge does in conventional electronics.[1] Ferromagnetic semiconductors, such as the thoroughly investigated (Ga,Mn)As, are especially promising as the materials for spintronics since they interrelate both semiconducting and magnetic properties.[2] One of the hot topics including those materials is controllable dynamics of individual magnetic domain wall (DW) in nanowires.[3,4] It is essential here that DW contributes an extra electrical resistance to a ferromagnetic wire.

We investigated (Ga,Mn)As-based nanostructures shaped like three narrow strips, or nanowires, (labeled A, B, and C) joined in one point, which form an angle of 120° between each close pair (Fig. 1a). In ferromagnetic nanowires, magnetic shape anisotropy, caused by the dipole interaction between magnetic moments, usually dominates over magneto-crystalline anisotropy caused by the crystal field and spin-orbit coupling. Magnetization direction is then forced to be oriented along the wire axis. Despite this is not generally true in the case of diluted ferromagnetic semiconductors, we actually assume it as well as magnetic equivalence of all three arms. Clearly, a region of inhomogeneous magnetization, or spin misalignment, has to appear at the junction, and we conveniently describe it in terms of the domain walls.

Magnetostatic energy of this system minimizes when the sense of magnetization vector, **M**, is conserved while going along any one of the three pairs of arms. Let this pair be BC, and the sense of **M** be from B to C. The vector **M** is then deflected by 60º at the junction, i.e. a 60º DW separates magnetic domains in the arms B and C. When the sense of **M** in the arm A is toward the junction, then magnetic domains in A and B are separated by a 120º DW, and those in A and C by a 60º DW. If, now, we reverse the magnetization direction in the arm A, the DWs between relevant arms counterchange. In each case the system contains two 60º DWs (with slight spin misaligned) and one 120º DW (with considerable spin misaligned). Any other non-equivalent magnetic configuration is energetically less favorable as it contains three 120º DWs.

Our structure represents a three terminal device in which an electric current can be driven through any of the three pairs of arms. In each case, charge carriers have to cross either 60º DW or 120º DW at the junction surmounting then an extra resistance associated with the DW.[5,6] Generally, different effects contribute to the DW resistance in a given material. Some of them follow basically from the Lorentz force,[7] others are related to the spin-dependent scattering of charge carriers in the wall.[6,8] Interestingly, the DW resistance can also be negative. It occurs in a case when the dominant effect of DW consists in the local destruction of quantum-interference contribution to the resistivity caused by the effect of weak-localization.[9,10] Here, we assume only that the DW resistance rises with the degree of spin misalignment in a wall.

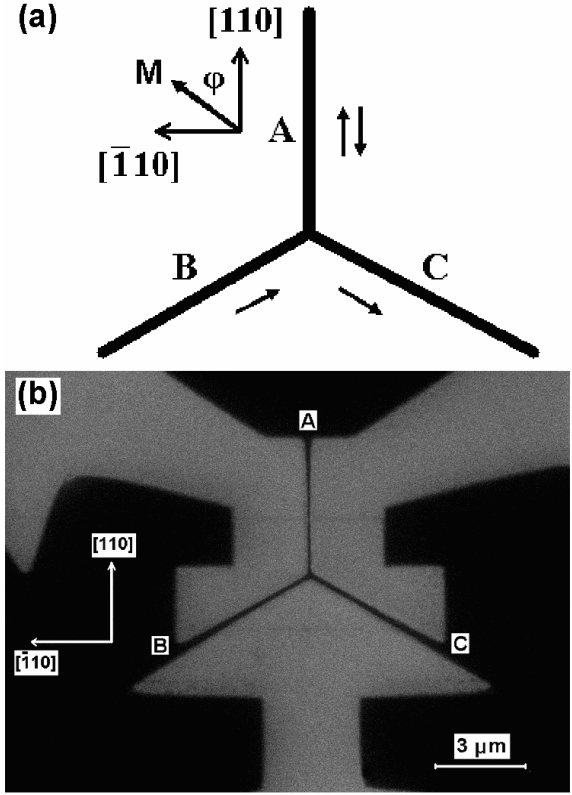

FIG. 1 Three-arm nanostructure: (a) Schematic view where arrows represent magnetization vectors, (b) scanning electron microscope image where darker contrast corresponds to the non-etched conducting areas of the structure.

We fabricated the three-arm structure (TAS) from monocrystalline $Ga_{0.96}Mn_{0.04}As$ layer, 15 nm thick, grown by a low-temperature molecular-beam epitaxy method on the (001) face of semi-insulating GaAs substrate.[11] In this layer, the ferromagnetic ordering of magnetic moments of Mn ions occurs below 60 K and is mediated by holes that give rise to the metallic-type conductivity. A lattice mismatch on the interface introduces to such-grown layers a compressive strain, which makes that an easy axis of magnetization lies in plane of the layer. In addition, (Ga,Mn)As layers display a complex in-plane magneto-crystalline anisotropy, depending on the hole concentration and temperature.[12-14] There are two equivalent easy axes of magnetization along two in-plane $\langle 100 \rangle$ directions (cubic anisotropy). Another competitive easy axis is along either [110] or [$\bar{1}$10] direction (uniaxial anisotropy). Hamaya et al.[15] determined all the lowest-order anisotropy constants, i.e. cubic, uniaxial (along the [110] direction) and shape, in 0.8 μm-wide wire tailored from $Ga_{0.962}Mn_{0.038}As$ layer having physical parameters very similar to those of our layers. We use these constants in further discussion of our results.

We have designed the TAS in a configuration where the arm A is along the [110] easy direction. The arms B and C, are oriented closely to the cubic easy axes. All the arms are about 200 nm wide and 3 to 5 μm long. The structures were fabricated using electron-beam lithography pattering and chemical etching (Fig. 1b). The arm terminals were supplied with Ohmic contacts. We measured resistance, R, between each pair of arms at T = 4.2 K as a function of a magnetic field, H, applied parallel to the arm A.

In strong magnetic fields, resistance of each pair of arms decreases with increasing field. That negative magnetoresistance (MR) is a common property of ferromagnetic (Ga,Mn)As layers,[2,16] which is generally understood as a reduction of spin-disorder scattering of charge carriers caused by ordering of Mn spins in an external magnetic field. Competitive mechanism, dominating at the lowest temperatures, is the magnetic-field-induced destruction of weak-localization. MR-features characteristic of individual samples appear only in a narrow range of magnetic field enclosing magnetization hysteresis. All the results presented here refer to that weak-field range, but one must remember that they are imposed on a pronounced negative-MR background.

The most striking result observed in the TAS is the symmetry of the R(H) curves, which appears for the AB and AC pairs of arms (Fig. 2). Reversal of the direction of magnetic field applied parallel to the arm A, is equivalent to the exchange of the pair AB into AC, and vice versa. All the R(H) curves display a hysteresis loop corresponding to the hysteresis loop of magnetization. It must be noticed a "remnant resistance", an analogue to the remnant magnetization, that appears in zero magnetic field. It means that resistance of either two pairs of arms can assume one of two stable values, depending on the previous magnetization direction in the arm A. Instead, the R(H) curves for the BC pair recorded at two opposite runs of a magnetic field applied parallel to the arm A, do not show any remnant resistance (Fig. 3).

Thus we have realized a bistable device, or a switch, in which momentarily applied magnetic field triggers transition from a lower to higher resistance in one pair of the arms, and the opposite transition in the other pair. It is worth noting that the resistance jumps at the transition (remnant resistances) are exactly the same, with an accuracy of the measurements, for the both pairs of arms, despite their total resistances differ by several percent; cf. Fig.2.

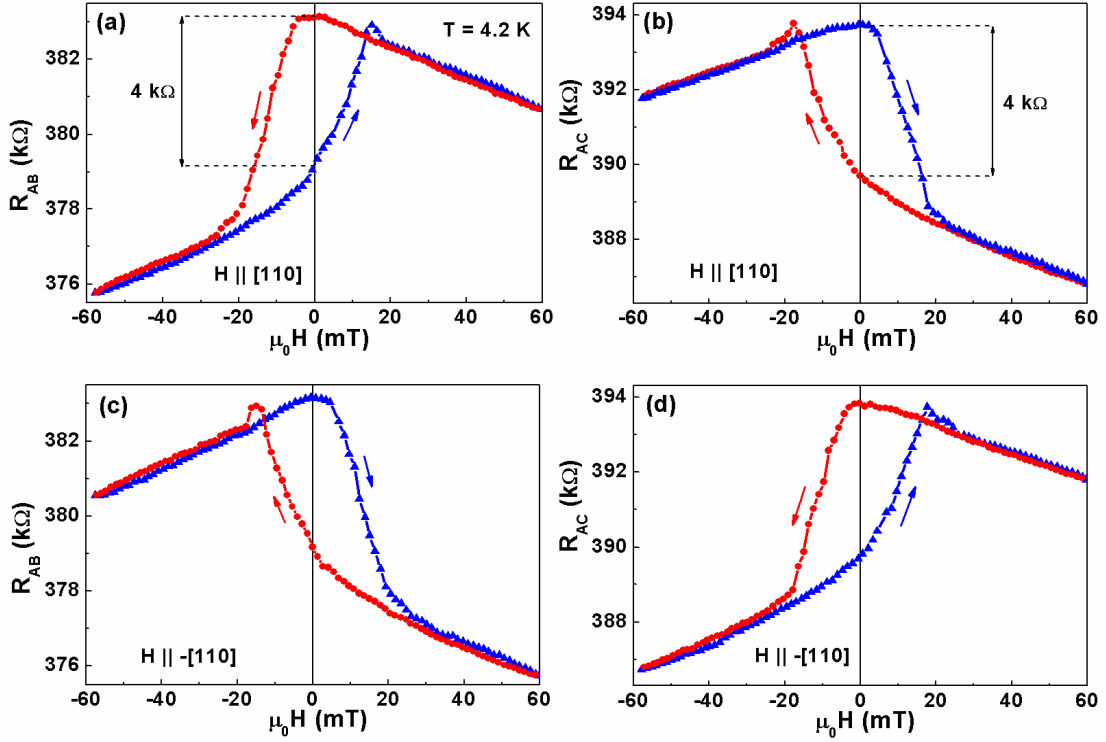

FIG. 2 Electrical resistances $R_{AB}$ (a), (c), and $R_{AC}$ (b), (d) measured between terminals of the respective arms AB and AC of the nanostructure as a function of a magnetic field oriented along the arm A swept in two opposite directions (differentiated by triangles or circles). The magnetic field direction in (c) and (d) was reversed with respect to the case (a) and (b). The magnitude of "remnant resistance" at zero magnetic field is marked. The resistance was measured applying a probing alternative voltage of about 3 mV.

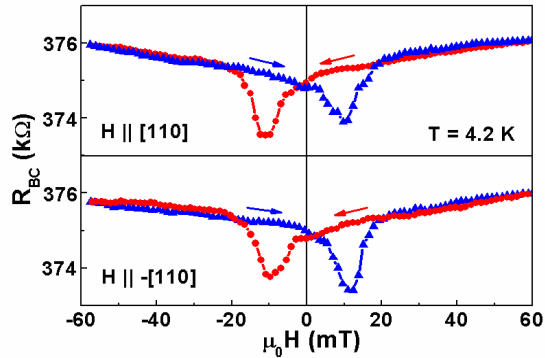

FIG. 3 Electrical resistance $R_{BC}$ measured between terminals of the arms B and C of the device as a function of a magnetic field oriented along the arm A swept in two opposite directions (differentiated by triangles or circles). The magnetic field direction in the lower plot was reversed with respect to the case of the upper plot. No "remnant resistance" appears.

To discuss the obtained results, we analyze configuration of spontaneous magnetization in the TAS in zero magnetic field. Magnetostatic energy in each arm of the structure is the sum of three anisotropy energies:

$$E = (K_C/4)\cos^2 2\varphi + K_U \sin^2 \varphi + K_S f(\varphi), \qquad (1)$$

where $K_C$, $K_U$, and $K_S$ are the in-plane cubic, uniaxial and shape anisotropy constants, respectively. $\varphi$ is the angle between the [110] direction and the magnetization vector, and $f(\varphi)$ stands for $\sin^2 \varphi$, $\cos^2(\varphi - 30°)$ and $\cos^2(\varphi + 30°)$ for the arms A, B and C, respectively. Minimizing $E$ with respect to $\varphi$, i.e. demanding $\partial E/\partial \varphi = 0$ and $\partial^2 E/\partial \varphi^2 > 0$, we determine stable magnetization direction in each arm. Using the anisotropy constants[15] (in erg/cm$^3$): $K_C$ = 3500, $K_U$ = 1800 and $K_S$ = 3000, we get $\varphi$ = 0º for the arm A, and $\varphi$ = ±136.4º for the arms B and C, respectively. Thus we find that the spontaneous magnetization is exactly aligned with the arm A, and is oriented closely to the [100] and [010] easy axes in the arms B and C, respectively. Consequently, there is a ~92º DW between the arms B and C, and either ~44º DW between A and B and ~136º DW between A and C, or vice versa, depending on the sense of vector **M** in the arm A.

Thus the essence of the observed effect consists in that the reversal of magnetization direction in the arm A by an external magnetic field switches a higher-resistance DW from one current path to the other, i.e. from AB to AC, or vice versa, while the magnetization in the arms BC is locked. Accordingly, the remnant resistance represents the difference in magnitudes of the resistance of the ~136º DW and ~44º DW involved in this effect. Its value, counted for unit area of DW, is of about 12 Ωμm$^2$.

In conclusion, we have for the first time demonstrated an effect of hysteretic magnetoresistance in (Ga,Mn)As-based nanostructure, where zero-field resistance depends on the direction of previously applied magnetic field. The three-arm structure realizing this effect, is in essence a two-state device, basic non-volatile-memory element. It represents also a three-terminal device that has two complementary outputs, what means that when one pair of arms is in the high-resistance state another is in the low-resistance state, and vice versa.

One of the authors (J. S.) would like to acknowledge the support from the research project financed by European Aeronautic Defence and Space Company (EADS)